\documentclass[aps,twocolumn,english, superscriptaddress,eqsecnum,showpacs,longbibliography]{revtex4-1}
\usepackage{epsfig}
\usepackage{graphicx}
\usepackage{dcolumn}
\usepackage{bm}
\usepackage{multirow}

\usepackage{amssymb}
\usepackage{amsmath}
\usepackage[colorlinks,citecolor=blue,linkcolor=red,urlcolor=blue]{hyperref}
\usepackage{xcolor}

\usepackage[colorlinks,citecolor=blue,linkcolor=red,urlcolor=blue]{hyperref}

\begin{document}

\title{Short-imaginary-time quantum critical dynamics in the J-Q$_3$ spin chain}
\author{Yu-Rong Shu}
\affiliation{School of Physics and Materials Science, Guangzhou University, Guangzhou 510006, China}
\affiliation{Research Center for Advanced Information Materials, Guangzhou University, Guangzhou 510006, China}
\author{Shuai Yin}
\email{yinsh6@mail.sysu.edu.cn}
\affiliation{School of Physics, Sun Yat-Sen University, Guangzhou 510275, China}
\date{\today}

\begin{abstract}
  We study the short-imaginary-time quantum critical dynamics (SITQCD) in the J-Q$_3$ spin chain, which hosts a quasi-long-range-order phase to a valence bond solid transition. By using the scaling form of the SITQCD with a saturated ordered phase, we are able to locate the critical point at $q_{\rm c}=0.170(14)$. We also obtain the critical initial slip exponent $\theta=-0.507(3)$ and the static exponent $\beta/\nu=0.498(2)$. More strikingly, we find that the scaling dimension of the initial order parameter $x_{0}$ is close to zero, which suggests that the initial order parameter is a marginal operator. As a result, there is no initial increase behavior of the order parameter in the short-imaginary-time relaxation process for this model, which is very different from the relaxation dynamics in the Ising-type phase transitions. Our numerical results are realized by the projector quantum Monte Carlo algorithm.

\end{abstract}

\maketitle

\section{Introduction}\label{sec:intro}
Nonequilibrium dynamics of quantum phase transitions has been an attractive topic in condensed matter physics and statistical physics in recent decades~\cite{dziarmaga2010ap,polkovnikov2011rmp}. Among different types of nonequilibrium dynamics, the quantum imaginary-time relaxation stands out as a usual method to find the ground state of quantum many-body systems. Moreover, algorithms based on the imaginary-time relaxation have been designed for quantum computations recently~\cite{love2020natphy,motta2020natphy}. Besides, studies on the imaginary-time evolution also reap great benefits~\cite{grandi2011prb,grandi2013jpcm,liu2013prb,avdoshkin2019,beach2019prb}. For example, it has been shown that in the driven critical dynamics, the imaginary- and real-time dynamics bare considerable similarities~\cite{grandi2011prb}, but the former is much easier to be realized numerically, especially for systems in higher dimension~\cite{sandvik2010aip,liu2013prb,grandi2011prb,grandi2013jpcm}. In addition, by comparing with the short-time critical dynamics in classical dissipative systems~\cite{janssen1989zpb,huse1989prb,albano2011rpp}, the scaling theory for the short-imaginary-time quantum critical dynamics (SITQCD) has been developed~\cite{yin2014prb,zhang2014pre} by analogy with its classical counterparts~\cite{lizb1994jpa,lizb1995prl,lizb1996pre,zheng1996prl,zheng1998ijmpb}. This theory provides efficient methods to determine the critical properties in the short-time region, overcoming the difficulties induced by the critical slowing down~\cite{yin2014prb,zhang2014pre,shu2017prb}.

In imaginary-time evolution, the system is controlled by low-lying energy levels so that universal power-law behaviors can exist during the evolution after a transient microscopic time~\cite{sachdev1999,yin2014prb,zhang2014pre}. In the Ising-type phase transition, the SITQCD theory shows that the critical initial slip of the order parameter $D(\tau)\propto D_{0}\tau^{\theta}$ exists when an initial state with small $D_{0}$ and zero correlation is prepared. Therein $\theta$ is the critical initial slip exponent and is positive for the quantum Ising model in both one and two dimension~\cite{yin2014prb,zhang2014pre,shu2017prb}. Namely, with a small initial value $D_{0}$, in early stage of the evolution, $D(\tau)$ does not decrease towards its ground-state value $0$. Instead, $D(\tau)$ counterintuitively experiences an increasing stage, which scales as $\tau_{\rm cr}\sim D_{0}^{-z/x_{0}}$ with $x_0$ being the scaling dimension of $D_{0}$~\cite{yin2014prb,zhang2014pre,shu2017prb}. For the quantum Ising model in both one and two dimension, $x_0$ is positive and $D_0$ is a relevant scaling variable, resulting the initial increase of $D(\tau)$.

Here, we study the SITQCD of the one-dimensional ($1$D) J-Q$_{3}$ model by means of quantum Monte Carlo (QMC) simulations. The Hamiltonian of the J-Q$_{3}$ chain is given by~\cite{sandvik2007prl,tang2011prl}
\begin{equation}
\label{eq:hamiltonian}
H=-J\sum_{i=1}^{L}P_{i,i+1}-Q\sum_{i=1}^{L}{P_{i,i+1}P_{i+2,i+3}P_{i+4,i+5}},
\end{equation}
where $J$ and $Q$ are both antiferromagnetic (AF) couplings and $P_{i,i+1}$ denotes the two-spin singlet operator
\begin{equation}
\label{eq:pij}
P_{i,i+1}=\frac{1}{4}-{\bf S}_{i}\cdot{\bf S}_{i+1}.
\end{equation}
The standard $J$ interactions tend to form the quasi-long-range-order (QLRO) phase that is in the class of the standard critical Heisenberg chain, while the multi-spin $Q$ terms favor a doubly-degenerate valence bond solid (VBS) phase. A transition appears at $q_{\rm c}=(Q/J)_{\rm c}\approx 0.16$~\cite{tang2011prl,sanyal2011prb}, separating the QLRO phase from the VBS phase. The same kind of phase transition also occurs in the well-studied J$_1$-J$_2$ spin chain~\cite{okamoto1992pla,eggert1996prb} at the coupling ratio $J_{2}/J_{1}=0.241167(5)$~\cite{eggert1996prb}. However, due to the ``sign problem'' caused by the next-nearest-neighbor frustrating J$_{2}$ interactions, QMC simulations of the J$_1$-J$_2$ model is hardly available. In addition, an akin J-Q$_{2}$ chain in the same J-Q family also has similar properties, but the VBS order is weaker in it~\cite{sandvik2007prl,tang2011prl}. In two dimension, the J-Q model exhibits a fascinating deconfined quantum phase transition between the N{\'e}el and VBS phase~\cite{sandvik2007prl}.

The rest of the paper is organized as follows. In Sec.~\ref{sec:stcd_theory}, we will review the SITQCD theory and the scaling relations that is useful in our study. The QMC method employed in this work will be outlined in Sec.~\ref{sec:methods}. We will present our numerical results in Sec.~\ref{sec:res} and discuss our findings in Sec.~\ref{sec:discussion}. A summary given in Sec.~\ref{sec:conclusion}.


\section{Short-imaginary-time quantum critical dynamics scaling theory}
\label{sec:stcd_theory}
For a quantum state $|\Psi(\tau)\rangle$, the imaginary-time evolution of the wave function is described by the imaginary-time Schr{\"o}dinger equation~\cite{justin1996,altland2006}. Near the critical point, $|\Psi(\tau)\rangle$ is governed by the low-energy levels during the imaginary-time evolution as the high energy levels decay very fast. According to the theory of SITQCD, observable $\mathcal{O}$ should obey the following scaling form~\cite{yin2014prb,zhang2014pre}
\begin{equation}
\label{eq:scaling_hypo}
\mathcal{O}(\tau,g,D_0,L)=b^{\phi}\mathcal{O}(\tau b^{-z} ,gb^{\frac{1}{\nu}},D_{0}b^{x_0} ,Lb^{-1}),
\end{equation}
in which $\tau$, $g$, $D_0$ and $L$ represent the imaginary time, the distance to the critical point, the initial value of the order parameter, and the system size, respectively. $z$ is the dynamic exponent, and $\nu$ is the correlation length exponent. $x_0$ is the dimension of $D_{0}$, and $\phi$ is related to the quantity $\mathcal{O}$ studied. For instance, $\phi=-\beta/\nu$ (with $\beta$ being the order parameter exponent) for the order parameter and $\phi=0$ for the dimensionless variable. There are two ``\textit{apparent}" fixed points that can be readily identified for $D_{0}$: One is $D_0=0$, the other is $D_0=D_{\rm sat}$ with $D_{\rm sat}$ being the maximum value of $D$ (the saturated value, which depends on the model studied). $D_{0}=0$ and $D_{0}=D_{\rm sat}$ represent completely disordered and ordered states, respectively, which do not change under scale transformation. Moreover, these two fixed points do not depend on the scaling dimension of $D_{0}$.

By choosing the scaling factor $b=\tau^{\frac{1}{z}}$, one obtains the scaling form of $\mathcal{O}$,
\begin{equation}
\mathcal{O}(\tau,g,D_0,L^{-1})=\tau^{\frac{\phi}{z}}f_{\mathcal{O}}(g\tau^{\frac{1}{\nu z}}, D_{0}\tau^{\frac{x_0}{z}}, L^{-1}\tau^{\frac{1}{z}}),
\label{eq:scalform}
\end{equation}
in which $f_{\mathcal{O}}$ is the scaling function related to $\mathcal{O}$.
For small $D_0$, in the short-time region, $f_{\mathcal{O}}$ can be expanded as a series of $D_{0}\tau^{\frac{x_0}{z}}$.
Note that the correlation length $\xi$ of initial state has to very short as required by the SITQCD theory~\cite{janssen1989zpb,yin2014prb}. With $\xi\rightarrow 0$, the derivatives of the free energy are analytic. Besides, in the short-time region, $f_{\mathcal{O}}$ is a continuous function of $D_{0}\tau^{\frac{x_0}{z}}$, so that one can perform series expansion of $f_{\mathcal{O}}$ in terms of $D_{0}\tau^{\frac{x_0}{z}}$. Such treatment has proven to be valid in both classical short-time critical dynamics~\cite{janssen1989zpb} and the SITQCD theory~\cite{yin2014prb} already.
Take the order parameter $D$ for an example, the leading part of the scaling form obeys
\begin{equation}
D(\tau,g,D_0,L^{-1})=D_0\tau^{\theta}f_{D}(g\tau^{\frac{1}{\nu z}},L^{-1}\tau^{\frac{1}{z}}),
\label{eq:scalform1}
\end{equation}
in which the critical initial slip exponent $\theta$ reads
\begin{equation}
\label{eq:xotheta}
\theta=\frac{x_0}{z}-\frac{\beta}{\nu z}.
\end{equation}
When $\theta>0$, the order parameter increases in the initial stage of the evolution. This is the case for the quantum Ising model in both one and two dimension~\cite{yin2014prb,shu2017prb}. Therein the initial order parameter is relevant and $x_0$ is larger than $\beta/\nu$.

However, when the initial order parameter is marginal, i.e. $x_0=0$, Eq.~(\ref{eq:xotheta}) gives $\theta=-\frac{\beta}{\nu z}$. In this situation, the order parameter will not increase with $\tau$. Instead, it will decay as $D\sim D_0\tau^{-\frac{\beta}{\nu z}}$, similar to its long-time relaxation. We will find that this is just the case for the J-Q$_{3}$ spin chain~(\ref{eq:hamiltonian}) studied here.

Moreover, when the initial order parameter $D_0$ is chosen at its apparent fixed points, i.e., $D_0=0$ or $D_0=D_{\rm sat}$, Eq.~(\ref{eq:scaling_hypo}) shows that the $k$-th moment of the order parameter with $D_{0}$ being at its fixed point satisfies
\begin{equation}
\label{eq:fDk}
D^{k}(\tau,L^{-1})=\tau^{-k\frac{\beta}{\nu z}}f_{D^{k},D_0}(g\tau^{\frac{1}{\nu z}},L^{-1}\tau^\frac{1}{z}).
\end{equation}

Besides the order parameter, the SITQCD behavior also manifests itself in the imaginary-time correlation function of $D$~\cite{huse1989prb,tome1998pre,shu2017prb}
\begin{equation}
\label{eq:ctau_1}
C(\tau)=\lim_{D_{0}\rightarrow 0}{\frac{D(\tau)}{D_{0}}}=L\langle{\hat{D}(0)\hat{D}(\tau)\rangle},
\end{equation}
in which $\hat{D}$ is the operator of the dimer order parameter at imaginary time $0$ and $\tau$. $\langle \cdots\rangle$ represents statistical average of the operators.
It has been shown that $C(\tau)$ satisfies $C(\tau)\propto \tau^{\theta}$ in the thermodynamic limit, while for finite-size systems, the scaling form of $C(\tau)$ at the critical point is~\cite{huse1989prb,tome1998pre,shu2017prb}
\begin{equation}
\label{eq:ctau_2}
C(\tau,L)=\tau^{\theta}f_{C}(\tau L^{-z}).
\end{equation}
According to Eq.~(\ref{eq:ctau_2}), when the initial order parameter is marginal, $C(\tau)$ decays as $C(\tau)\sim \tau^{-\frac{\beta}{\nu z}}$, as will be seen in the J-Q$_{3}$ chain.



The scaling theory of the SITQCD can be employed to determine the critical properties~\cite{yin2014prb,zhang2014pre,shu2017prb}. For example, to determine the critical point, the initial order parameter $D_0$ can be chosen as its fixed values to lessen the variables in Eq.~(\ref{eq:scalform}). In this situation, the dimensionless variable, such as the average sign of the order parameter $I(\tau)$, defined as $I(\tau)=\langle{{{\rm sgn}\{D(\tau)\}}}\rangle$~\cite{oliveira1992epl,soares1997prb}, satisfies
\begin{equation}
\label{eq:scaling_itau}
I(\tau,g)=f_{I}(\tau L^{-z},L^{\frac{1}{\nu}}g).
\end{equation}
For a fixed aspect ratio $\tau L^{-z}$, Eq.~(\ref{eq:scaling_itau}) shows that $I(\tau,g)$ cross at $g=0$ for different system sizes. Accordingly, the critical point can be determined. In addition, by using Eq.~(\ref{eq:fDk}) at $g=0$, one can determine the static exponent $\beta/\nu$. Moreover, $\theta$ can be estimated from Eq.~(\ref{eq:scalform1}) and Eq.~(\ref{eq:ctau_2}). For the case where $D_{0}$ is relevant, Eq.~(\ref{eq:ctau_2}) is simpler in practice as it takes the limit $D_{0}\rightarrow 0$ in advance.


\section{Numerical method}
\label{sec:methods}
In this section, we will introduce the QMC method used in our calculations briefly. The projector QMC method employed in this work is based on the stochastic series expansion (SSE) QMC method~\cite{sandvik2010aip}.

In imaginary time, the Schr{\"o}dinger equation describes the evolution of a quantum state $|\Psi(\tau)\rangle$ as~\cite{altland2006,justin1996}
\begin{equation}
  \label{eq:s_eq}
  \partial_{\tau}|\Psi(\tau)\rangle =-H|\Psi(\tau)\rangle.
\end{equation}
A formal solution of the Schr{\"o}dinger equation is given by
\begin{equation}
  \label{eq:formal}
  |\Psi(\tau)\rangle=U(\tau)|\Psi(\tau_{0})\rangle,
\end{equation}
in which $U(\tau)={\rm e}^{-\tau H}$ is the imaginary-time evolution operator and $\tau_{0}$ is the starting time of the evolution. The expectation value of an operator $\mathcal{\hat{O}}$ at $\tau$ is then
\begin{equation}
  \label{eq:expectation}
  \mathcal{O}(\tau)=\frac{1}{Z}\langle\Psi(\tau)|\mathcal{\hat{O}}|\Psi(\tau)\rangle,
\end{equation}
where the normalization is defined as
\begin{equation}
  \label{eq:formal}
  Z=\langle\Psi(\tau)|\Psi(\tau)\rangle=\langle\Psi(\tau_{0})|{\rm e}^{-\tau H}{\rm e}^{-\tau H}|\Psi(\tau_{0})\rangle.
\end{equation}
The central idea of the projector QMC method is to perform series expansion of $U(\tau)$ in the normalization
\begin{equation}
  \label{eq:expand}
  Z=\sum_{n}^{\infty}\sum_{S_{n}}{\langle\Psi(\tau_{0})|\frac{\beta^{n}}{n!}S_{n}|\Psi(\tau_{0})\rangle},
\end{equation}
with $S_{n}$ denoting the operator sequence and $\beta=2\tau$. The expansion order $n$ can be truncated to some maximum length that causes no detectable error.
The operator sequence and states are then importance-sampled and measurements can be done accordingly. To gain efficiency, we employ a global loop-update scheme in the importance sampling procedure~\cite{sandvik2010aip,farhi2012pra}. In our calculations, we perform $10^{5}$ equilibration steps followed by at least $100$ bins of successive measurements, each with $10^{5}$ Monte Carlo steps, in order to ensure statistical errors are under control.

Comparing with the SSE method, in the projector method, the imaginary-time axis can have different or fixed boundary states, which is actually crucial for realizations of different initial states in this study. Besides, for short evolution times, a binomial weight factor should also be inserted in Eq.~(\ref{eq:expand}) in order to obtain accurate expectation values as different propagated states has different contributions when $\tau$ is not large. At long times, the effect of the weight factor becomes negligible and the measurements can be done in the ``middle'' of the projection axis far away from the boundaries~\cite{farhi2012pra}.

In addition, in the projector QMC method, apart from the standard $S^{z}$ basis, the valence bond basis can also be applied~\cite{tang2011prl,beach2012npb}. Here, we consider different initial states, including VBS, AF and disordered states. The valence bond basis has $S^{z}_{\rm tot}=0$ so that it is convenient in realizing VBS states. For disordered/AF states, the standard $S^{z}$ basis is more useful. Therefore, in our calculations, different basis will be used according to the initial state. Both the SSE and projector QMC method are well-documented and here we refer details of the methods to the literature~\cite{sandvik2010aip,farhi2012pra,beach2012npb}.

\section{Numerical Results}
\label{sec:res}
In this section, we present QMC results of the SITQCD in the QLRO-VBS transition of the J-Q$_{3}$ chain. First we will locate the critical point of the transition and then compute the critical initial slip exponent $\theta$. The static exponent ratio $\beta/\nu$ is then determined. By comparing $\theta$ and $\beta/\nu$, we find that their absolute value are almost equal to each other, namely $x_{0}$ very close to $0$, indicating a marginal $D_0$. The dynamical exponent $z$ of the J-Q$_{3}$ chain is known as $z=1$~\cite{sandvik2010aip}, which will be set as input.

In the J-Q$_{3}$ chain, the order parameter for the dimer order is defined as $D=(\sum_{i}^{L}{(-1)^{i}{\bf S}_{i}\cdot{\bf S}_{i+1}})/L$ or its $z$-component $D_{z}$. In the following, to keep simplicity, the full dimer order parameter and its $z$-component are both denoted as $D$.

\subsection{Determination of the critical point}
\begin{figure}[htbp]
  \includegraphics[width=\columnwidth,clip]{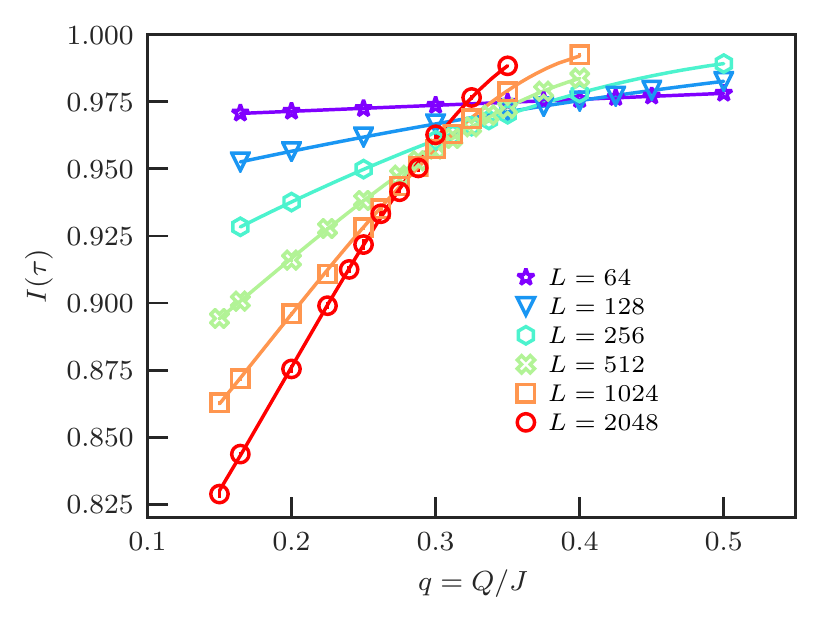}
  \vskip-3mm
  \caption{$I(\tau)$ for different coupling ratio $q=Q/J$ with sizes from $L=64$ to $2048$ at $\tau L^{-z}=1/16$. The errorbars are much smaller than the symbols (so do other figures in the following). The solid lines are polynomial fits to the data, up to cubic terms. }
  \label{fig:itau}
\end{figure}

To locate the critical point, the system is prepared in the VBS initial state, and then relaxes in the imaginary time. Here $D_{\rm sat}=3/8$ (full order parameter). We compute $I(\tau)$ for $L=48$ to $2560$ with a fixed aspect ratio $\tau L^{-z}=1/16$. In Fig.~\ref{fig:itau}, we plot $I(\tau)$ for $L=64$ to $2048$ to show how the crossing point of $L$ and $2L$ evolves with the increase of $L$.
The values of $I(\tau)$ are close to $1$ for all coupling ratios $q$, indicating that the system remain mostly in the VBS phase. It is obvious that the evolution time $\tau=L^{z}/16$ is too short for the system to get rid of the remanence of the initial VBS state.

\begin{figure}[htbp]
  \includegraphics[width=\columnwidth,clip]{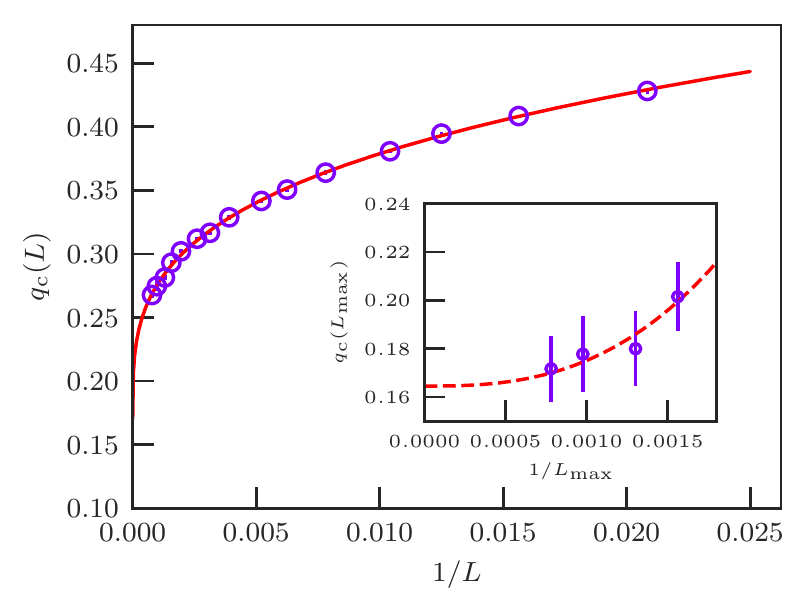}
  \vskip-3mm
  \caption{Main panel: dependence on system size of the crossing point of $I(\tau)$ for $L$ and $2L$. The solid line is a fit with the form of $q_{\rm c}(L)=q_{\rm c}+aL^{-\omega}$ to extract $q_{\rm c}$ in the thermodynamic limit. $q_{\rm c}$ is given by $0.170(14)$, with $a=0.81(3)$, $\omega=0.29(3)$ and $\chi^{2}$ per degree of freedom is $1.04$. Inset: dependence of $q_{\rm c}$ on the largest system size $L_{\rm max}$ included in the fitting. The dashed line is guide to eyes.}
  \label{fig:gcL}
\end{figure}
Using polynomials up to cubic terms to fit the data, we can extract the crossing point $q_{\rm c}$ of $I(\tau)$ for $L$ and $2L$. The dependence of $q_{\rm c}$ on the system size $L$ is shown in Fig.~\ref{fig:gcL}. Unlike usual cases where $q_{\rm c}(L)$ converges rapidly as $L$ increases, here $q_{\rm c}(L)$ exhibits a convex behavior, which suggests that the size effect in $q_{\rm c}$ is not negligible even at the largest-size system accessed. We use the form $q_{\rm c}(L)=q_{\rm c}+aL^{-\omega}$~\cite{binder1981prl} to fit $q_{\rm c}(L)$ and find that in the limit of $L\rightarrow \infty$, $q_{c}$ is $0.170(14)$, which agrees with an exact diagonalization (ED) result $q_{\rm c}=0.16478(5)$ given in a recent study~\cite{yang2020}. 

In the inset of Fig.~\ref{fig:gcL}, we show the dependence of $q_{\rm c}$ on the fitting range by changing the largest system size $L_{\rm max}$ included in the fitting.
As $L_{\rm max}$ increases, $q_{\rm c}$ approaches the ED result $q_{\rm c}=0.16478(5)$ rapidly.
In Ref.~\cite{yang2020}, the authors also use equilibrium QMC technique to extract the critical point $q_{\rm c}=0.21(4)$. In addition, our estimation of $q_{\rm c}$ has approximately the same error level with the equilibrium QMC result in Ref.~\cite{yang2020}. However, since the accessible system size (up to $L=256$) is much smaller compared to our result, it is possible that the equilibrium QMC study has not reach the region where the size effect in $q_{\rm c}(L)$ becomes clear.
Even though our result of $q_{\rm c}$ comes with large errorbar, the non-converging convex behavior of $q_{\rm c}(L)$ and slow decay of $I(\tau)$ on the QLRO side help to explain the reason why it is difficult for QMC studies (either equilibrium or nonequilibrium) to extract the precise critical point. Certainly, our result can be improved by accessing larger system sizes and data of better quality, which will consume much more computational resources and we will leave it to further studies. Since our estimation of $q_{\rm c}=0.170(14)$ only has moderate precision, we will use the ED estimation $q_{\rm c}=0.16478$~\cite{yang2020} in the following.

In Fig.~\ref{fig:gcL}, the aspect ratio $\tau L^{-z}$ is fixed at $1/16$ but we have also tried different values of the aspect ratio (data not shown). For larger $\tau L^{-z}$, the curve of $q_{\rm c}(L)$ is moving downwards but also becoming flatter, comparing to the one shown here, which makes it more difficult to analyze the size effect. In addition, as $\tau L^{-z}$ increasing towards $1$, the behavior of $q_{\rm c}(L)$ converges to ground-state results, requiring much more computational resources.
However, this does not mean the smaller $\tau L^{-z}$ is, the better. For small values, for instance $\tau L^{-z}=1/100$, the size required to reach the same scale of $\tau$ can be too large to simulate, since $\tau$ should also exceed the microscopic time $\tau_{\rm mic}$ so as not to fall in the non-universal stage. Therefore, it is better to choose a medium $\tau L^{-z}$ based on the consideration of balancing the shape of $q_{\rm c}(L)$, the system size available and simulation time. Even so, the SITQCD can still save a large amount of computation efforts.

\subsection{Determination of the exponent $\theta$}

\begin{figure}[htbp]
  \includegraphics[width=\columnwidth,clip]{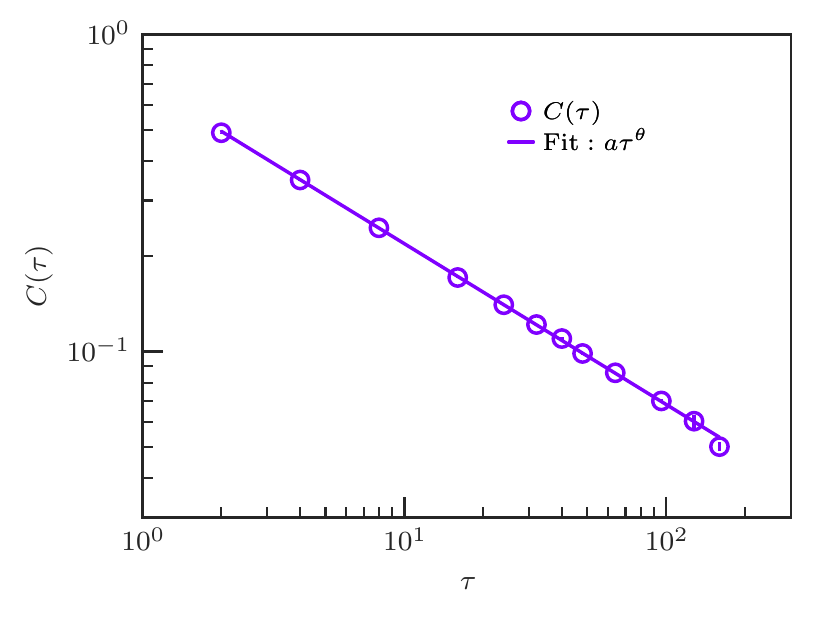}
  \vskip-3mm
  \caption{Dependence of $C(\tau)$ on the evolution imaginary time $\tau$ with fixed $\tau L^{-z}=1/16$. Power-law fitting shows the critical initial slip exponent $\theta=-0.507(3)$ with a prefactor $a=0.704(6)$. Double logarithmic scales are used.}
  \label{fig:theta}
\end{figure}

In order to determine $\theta$, we compute the imaginary-time correlation $C(\tau)$ for different $L$ ranging from $32$ to $2560$ with $D_0=0$ according to Eq.~(\ref{eq:ctau_2}). The aspect ratio is fixed at $\tau L^{-z}=1/16$. As shown in Fig.~\ref{fig:theta}, $C(\tau)$ does not increase with $\tau$ in the J-Q$_{3}$ spin chain, in contrast to the case of the quantum Ising model~\cite{yin2014prb,shu2017prb}. Instead, it decays with $\tau$ as a power law $C(\tau)\sim \tau^{\theta}$ with
\begin{equation*}
  \theta=-0.507(3).
\end{equation*}

\begin{figure}[htbp]
  \includegraphics[width=\columnwidth,clip]{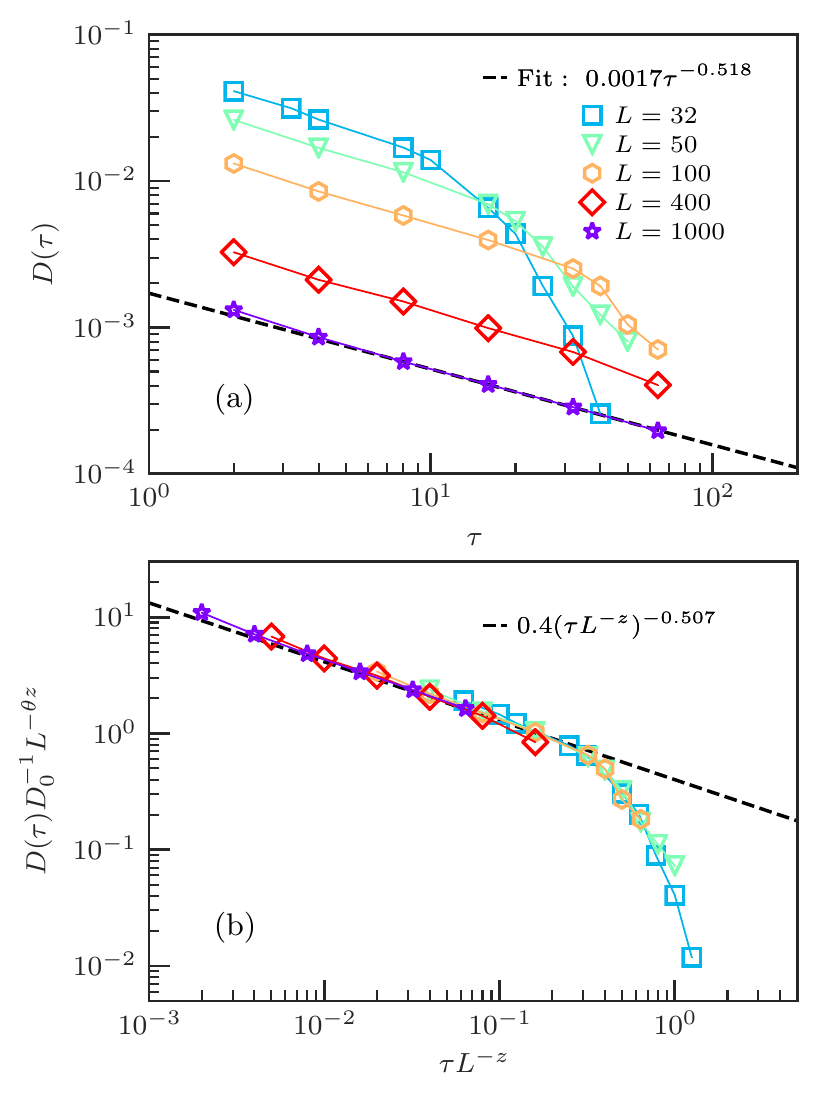}
  \vskip-3mm
  \caption{(a) Dependence of $D(\tau)$ on the evolution imaginary time $\tau$ for various sizes as marked. (b) Rescaled curves for (a) according to Eq.~(\ref{fss}). The dashed line in (a) is a power-law fit to show the exponent of $L=1000$ while the one in (b) is plotted to show the power-law behaviors of the rescaled curves.}
  \label{fig:theta1}
\end{figure}

To double check the exponent $\theta$ given by $C(\tau)$, we study the behavior of $D(\tau)$ when the initial state has non-zero but very small $D_{0}$, which is close to its apparent zero fixed point. For system of length $L$, the smallest positive value of $D$ is $1/L$ ($z$-component). This value is chosen as the initial $D_0$ for each size and the evolution of $D$ is shown in Fig.~\ref{fig:theta1}. In Fig.~\ref{fig:theta1} (a), it is clear that at the short-time stage, all $D(\tau)$ for various sizes satisfy a power law and the power-law range extends as $L$ increases. From Eq.~(\ref{eq:scalform1}), one finds that $D(\tau)\simeq D_0\tau^\theta f(0,0)+O(L^{-1}\tau^{1/z})$. Thus, $\theta$ can be fitted out by the short-time data of $D(\tau)$. We obtain $\theta$ as $\theta=-0.518(1)$ from the fitting of the data for $L=1000$. This value is close to the one obtained from $C(\tau)$ as we discussed above. The deviation between the two estimations may due to the finite-length of the $L=1000$ system, which is not large enough for $D(\tau)$ to get rid of finite-size effect as for systems of different size, $\theta$ drifts slightly.

Additionally, in Fig.~\ref{fig:theta1} (a), one finds that $D(\tau)$ drops in later times towards zero. The dropping time is earlier for system with smaller size. This demonstrates that the finite-size effects control the scaling in the late-time stage. Also, it means that the higher order terms of $L^{-1}\tau^{1/z}$ in the expansion of Eq.~(\ref{eq:scalform1}) dominate for large $\tau$ and small $L$. Moreover, for $g=0$, Eq.~(\ref{eq:scalform1}) is equivalent to
\begin{equation}
D(\tau,g,D_0,L)=D_0L^{\theta z}f_{DL}(L^{-z}\tau)
\label{fss}
\end{equation}
by the variable replacement. After rescaling $D(\tau)$ for different sizes according to Eq.~(\ref{fss}) with $\theta=-0.507$ as input, we find in Fig.~\ref{fig:theta1} (b) that all curves collapse onto each other. This result not only confirms the value of $\theta$, but also verifies Eq.~(\ref{fss}). Moreover, from Fig.~\ref{fig:theta1} (b), one finds that in the short-time region with small $\tau$, $f_{DL}(L^{-z}\tau)$ satisfies $f_{DL}(L^{-z}\tau)\propto(L^{-z}\tau)^\theta$, which recovers Eq.~(\ref{fss}) to $D(\tau)\propto D_0\tau^\theta$.


\subsection{Determination of the static exponent $\beta/\nu$}

Next, let us consider the static critical exponent $\beta/\nu$. As pointed out already, $D_{0}=D_{\rm sat}$ and $D_{0}=0$ are both apparent fixed points of Eq.~(\ref{eq:scaling_hypo}), giving the scaling form of Eq.~(\ref{eq:fDk}). Thus, we can estimate $\beta/\nu$ from these two different initial states here.

First we consider $D_0=D_{\rm sat}$.
Here, the calculations are performed in the valence bond basis with $D_{\rm sat}=3/8$.
It is obvious that $D(\tau)$, $D(\tau)^{2}$ should scale as $\tau^{-\beta/\nu z}$ and $\tau^{-2\beta/\nu z}$, respectively, for $g=0$ and a fixed $\tau L^{-z}$.
At longer times, $D(\tau)$ can be described using a power law. For $\tau=32-160$, the fitting gives $\beta/\nu=0.4919(2)$ along with a prefactor $a=0.383(1)$. For $\tau$ ranged from $96$ to $160$, we find
\begin{equation*}
\beta/\nu=0.498(2)
\end{equation*}
with $a=0.394(3)$. To on the safe side, the value of $0.498(2)$ is used as our final estimation of $\beta/\nu$. We will use this value to represent the asymptotic value of $\beta/\nu$.

\begin{figure}[htbp]
  \includegraphics[width=\columnwidth,clip]{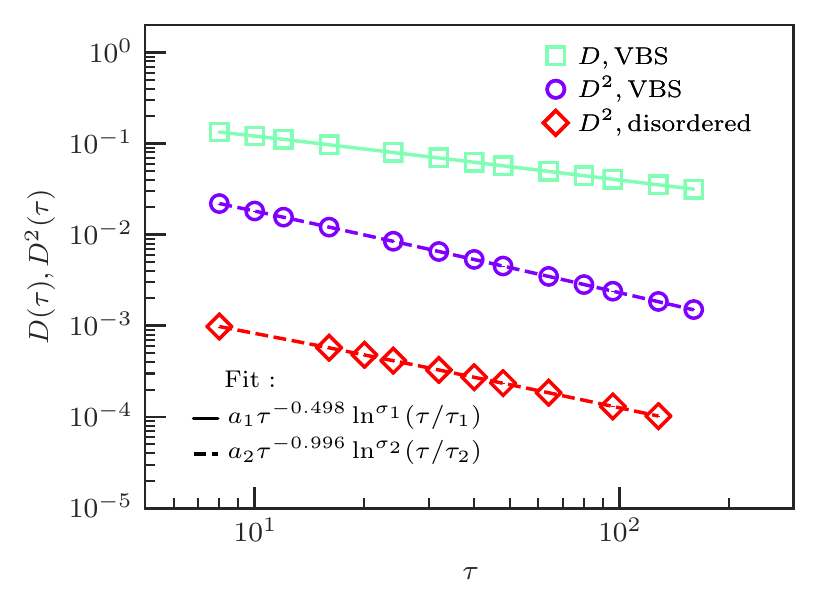}
  \vskip-3mm
  \caption{Power-law decay of $D(\tau)$ and $D^{2}(\tau)$ for different initial conditions, with power corresponds to $\beta/\nu$ and $2\beta/\nu$, respectively. Multiplicative logarithmic corrections to the power laws are included in order to obtain good fits. The finial estimation of $\beta/\nu$ is given by $0.498(2)$. }
  \label{fig:ordr}
\end{figure}

In order to include data of earlier times, by fixing $\beta/\nu= 0.498$, we consider a logarithmic correction in the fitting as
\begin{equation}
\label{eq:log}
D(\tau)=a_{1}\tau^{-\beta/\nu}\ln^{\sigma_{1}}(\tau/\tau_{1}).
\end{equation}
As shown in Fig.~\ref{fig:ordr}, we find that all data are well accounted for with the logarithmic correction. The fitting parameters are given by $a_{1}=0.3862(5)$, $\sigma_{1}=0.021(1)$ and $\tau_{1}=5.7(3)$. The logarithmic correction is actually not weak in this case.

Moreover, we observe similar behavior in $D^{2}(\tau)$ with $D_{0}=D_{\rm sat}$. With $\beta/\nu$ fixed at $0.498$, we use the form of $D^{2}(\tau)=a_{2}\tau^{-2\beta/\nu}\ln^{\sigma_{2}}(\tau/\tau_{2})$ to perform the fitting. We find that the curve is also well described but the logarithmic correction appears to be stronger in $D^{2}(\tau)$ with $a_{2}=0.136(4)$, $\sigma_{2}=0.34(2)$ and $\tau_{2}=1.0(1)$.

To further confirm the value of $\beta/\nu$, we consider the evolution starting from a disordered initial state with $D_{0}=0$. In this case $D(\tau)$ keeps zero and we study the behavior of $D^2(\tau)$. As seen in Fig.~\ref{fig:ordr}, the curve of $D^{2}(\tau)$ with $D_0=0$ is almost parallel to the corresponding curve with $D_0=D_{\rm sat}$, indicating identical critical exponents. By setting $\beta/\nu=0.498$ as input, we perform fitting using the same functional form and find out $a_{2}=0.0025(2)$, $\sigma_{2}=0.88(2)$ and $\tau_{2}=0.21(3)$ for $D^2(\tau)$ with $D_0=0$. The deviation between $D^{2}(\tau)$ in Fig.~\ref{fig:ordr} comes from the different definition of the order parameter (full component for $D_{\rm sat}$ and $z$-component for the disordered case).
In all cases, when including logarithmic term and allow the power $\beta/\nu$ to vary, the fittings give $\beta/\nu$ equals to  $0.511(6)$, $0.48(1)$ and $0.53(3)$ from $D(\tau)$ and $D^{2}(\tau)$ with $D_{0}=D_{\rm sat}$, $D^{2}(\tau)$ with $D_{0}=0$, respectively. These results are in agreement with $0.498(2)$ that extracted from the behavior of $D(\tau)$ at longer times.

Even though the origin of the logarithmic corrections is not totally clear to us, we hereby discuss the possible reasons of their presence.
In the fittings, these corrections are introduced in order to include the data at earlier time with $\beta/\nu$ fixed at the result extracted from the longer times, i.e. $\beta/\nu=0.498$.
However, in the short-imaginary-time scaling forms of $D(\tau)$ and $D^{2}(\tau)$, we did not consider the short-time corrections independently, like the finite-size corrections considered in equilibrium studies. Therefore, it is possible that the short-time scaling corrections are responsible for the presence of the logarithmic corrections.
Another possibility is the inaccurate estimate of the critical point.
In the field-theory description, the QLRO-VBS transition is driven by a marginal irrelevant operator. This marginal operator causes multiplicative logarithmic corrections in the QLRO phase but exactly at the critical point, the logarithmic correction should vanish~\cite{affleck1985prl,affleck1987prb}. However, as pointed out already, it is difficult to extract the exact critical point in our study. It is also possible $q_{\rm c}=0.16478$ that we taken from the ED study~\cite{yang2020} does not catch the exact critical point, thus causing the logarithmic corrections.


As mentioned above, the same kind of dimerization transition in this model also occurs in the frustrated J$_1$-J$_2$ spin chain.
In a recent work~\cite{mudry2019prb} on the $S=1/2$ J$_1$-J$_2$ XYZ chain, it is pointed out that for the isotropic J$_1$-J$_2$ spin chain, the dynamical exponent $z=1$ and the critical exponent $\eta$ should equal to $1$, which indicates that $\beta/\nu$ is $1/2$, agreeing with our estimation $\beta/\nu=0.498(2)$.
This consistency between our result and theirs not only confirms that the J-Q$_3$ spin chain~(\ref{eq:hamiltonian}) shares the same universality class with the J$_1$-J$_2$ spin chain, as pointed out previously~\cite{tang2011prl,sanyal2011prb,tang2015prb,patil2018prb,yang2020}, but also shows again the validity of the SITQCD method.
In addition, the QLRO-VBS transition in the J$_1$-J$_2$ chain is closely related to the spontaneous dimerization occurs in the spin-Peierls compound CuGeO$_3$~\cite{uhrig1998prb,weibe1999prb}. Our results of the J-Q$_3$ chain provide an alternative access to the same physics and inspire further experimental and computational explorations on the nature of the dimerization transition~\cite{tang2011prl,sanyal2011prb,tang2015prb,patil2018prb,yang2020}.

\subsection{$D_0$ as a marginal scaling variable}

By comparing the value of $\theta$ and $\beta/\nu$, we can find that their absolute values are very close to each other. According to Eq.~(\ref{eq:xotheta}), we infer that the initial order parameter $D_0$ is a marginal scaling variable with $x_0=0$. Under scale transformation in Eq.~(\ref{eq:scaling_hypo}), $D_0$ does not change. Accordingly, besides the two apparent fixed points, i.e., $D_0=0$ and $D_0=D_{\rm sat}$, all $D_0$ with zero initial correlation are fixed points of the transformation. As a result, Eq.~(\ref{eq:fDk}) should be applicable for all $D_0$ but with different scaling function $f_{D^{k},D_0}(gL^{\frac{1}{\nu}},\tau L^{-z})$.

\begin{figure}[htbp]
  \includegraphics[width=\columnwidth,clip]{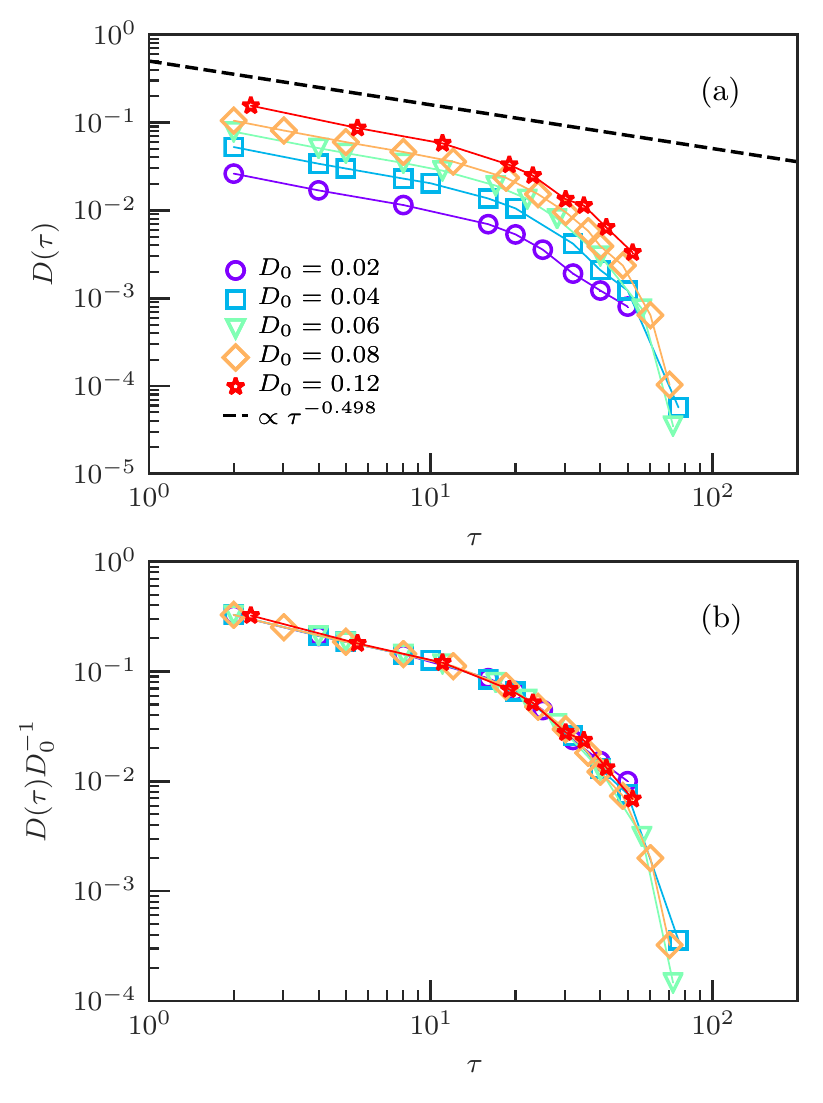}
  \vskip-3mm
  \caption{For a system with fixed $L=50$, curves of $D(\tau)$ versus $\tau$ before (a) and after (b) rescaling according to Eq.~(\ref{mard0}). For different $D_{0}$, $D(\tau)$ decays with almost the same exponent at earlier times, as indicated by the dashed line in (a).}
  \label{fig:fixl_d0}
\end{figure}

Here we argue that $f_{D^{1},D_0}(0,x)=D_{0}f_{D^{1}}(0,x)$ for any $D_0$. This equation is a direct generalization of Eq.~(\ref{eq:scalform1}).
Note that in Eq.~(\ref{eq:scalform1}), a small $D_{0}$ is required.
Since $\theta=-\beta/\nu z$ with $z=1$, Eq.~(\ref{eq:scalform1}) becomes $D(\tau)=D_{0}\tau^{-\beta/\nu z}f_{D}(0,x)$ for small $D_{0}$.
This scaling function is then identical with Eq.~(\ref{eq:fDk}) that is valid for $D_{0}=D_{\rm sat}$ since $D(\tau=0)=D_{0}$.
Namely, the relation $f_{D^{1},D_{0}}(0,x)=D_{0}f_{D^{1}}(0,x)$ is valid not only for small $D_{0}$ but also for the maximum $D_{\rm sat}$. Besides, in the short-time region, the scaling function is continuous in terms of $D_{0}$.
Therefore, one can conjecture that this relation should be valid for any value of $D_{0}$, such that
the evolution of $D(\tau)$ satisfies
\begin{equation}
D(\tau,L^{-1})=D_0\tau^{-\frac{\beta}{\nu z}}f(g\tau^{\frac{1}{\nu z}},L^{-1}\tau^\frac{1}{z}),
\label{mard0}
\end{equation}
in which the scaling function $f$ does not depend on $D_0$.

To examine Eq.~(\ref{mard0}), we consider the imaginary-time relaxation of $D(\tau)$ for various system sizes at $g=0$. In Fig.~\ref{fig:fixl_d0} (a), we find that $D(\tau)$ increases as $D_0$ increases. Moreover, for all $D_0$, in the short-time stage, $D(\tau)\propto \tau^{\theta}\sim \tau^{-\beta/\nu z}$. This indicates that for the purpose of extracting $\theta$ or $\beta/\nu$, $D_{0}$ does not necessarily restricted to small values in this situation.
In the late-time stage, the information contained in initial $D_0$ is ``forgotten'' and the curves for various $D_0$ tend to merge. In Fig.~\ref{fig:fixl_d0} (b), we rescale $D(\tau)$ with $D_0$ and find that all curves match with each other according to Eq.~(\ref{mard0}), showing that the scaling function $f$ does not depend on $D_0$ indeed.

\begin{figure}[htbp]
  \includegraphics[width=\columnwidth,clip]{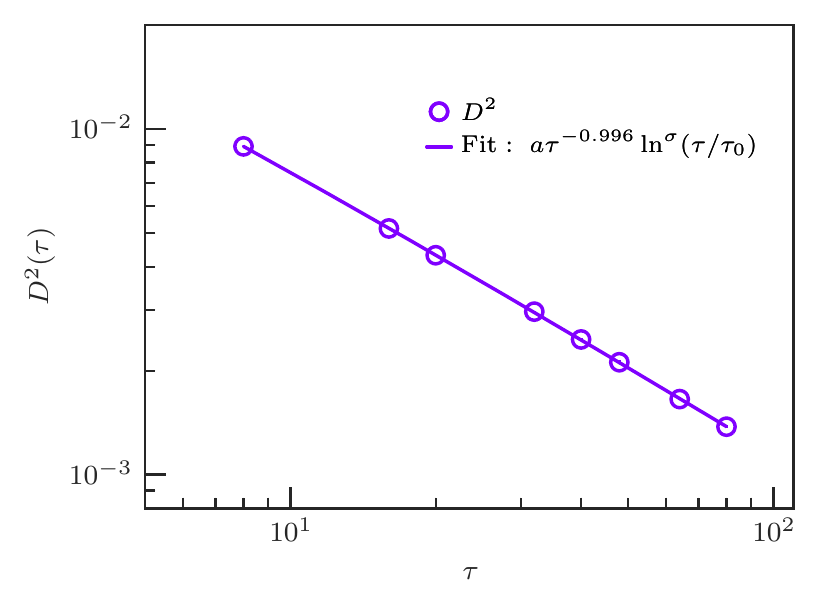}
  \vskip-3mm
  \caption{Dependence of $D^{2}$ on $\tau$ at the critical point relaxed from an AF starting state with $\tau L^{-z}=1/16$. $D^{2}(\tau)$ obeys Eq.~(\ref{mard0}) as well.}
  \label{fig:d2_afgc}
\end{figure}

Based on this, it is tempting to examine the behavior of $D(\tau)$ when the initial state has magnetic order.
We can infer that even the system is relaxed from an AF state with $D_{0}=0$, Eq.~(\ref{mard0}) should also be satisfied as long as the correlation length vanishes.
In Fig.~\ref{fig:d2_afgc}, we show the behavior of $D^{2}(\tau)$ instead of $D(\tau)$ as $D(\tau)$ is zero in this case. It is obvious that at the critical point, $D^{2}(\tau)\sim \tau^{-0.996}$, multiplied by a logarithmic correction with $a=0.10(4)$, $\tau_{0}=0.04(3)$ and $\sigma=1.2(2)$.
The behavior of $D^{2}(\tau)$ is very similar to the results with $D_{0}=D_{\rm sat}$ or $D_{0}=0$ shown in Fig.~\ref{fig:ordr}. Such result again reflects the marginal role of $D_{0}$ in the imaginary-time relaxation process.

\section{Discussion}
\label{sec:discussion}

Here we discuss the possible reasons for the marginal $D_0$. In the quantum Ising model, the positive $\theta$ is induced by the fact that the critical point is shifted down towards the ordered phase compared with its mean-field value. Thus the uncorrelated initial state ``feels" an ordered phase when the system is in the vicinity of the real critical point~\cite{yin2014prb}. In contrast, in the present case, the QLRO phase is a critical phase. Therefore, there is no proper mean-field solution for this model. In addition, the gap in the VBS phase is induced by a marginally relevant operator in the VBS phase from the field theory and this leads to the opening of an initially exponentially small gap~\cite{affleck1985prl,affleck1987prb}, in contrast to the Ising case that the gap is a power function of the distance to the critical point. These elements make the phase transition seems quite soft compared with the Ising case. The initial order parameter thus only plays a marginal role in the imaginary-time relaxation process.

\begin{figure}[htbp]
  \includegraphics[width=\columnwidth,clip]{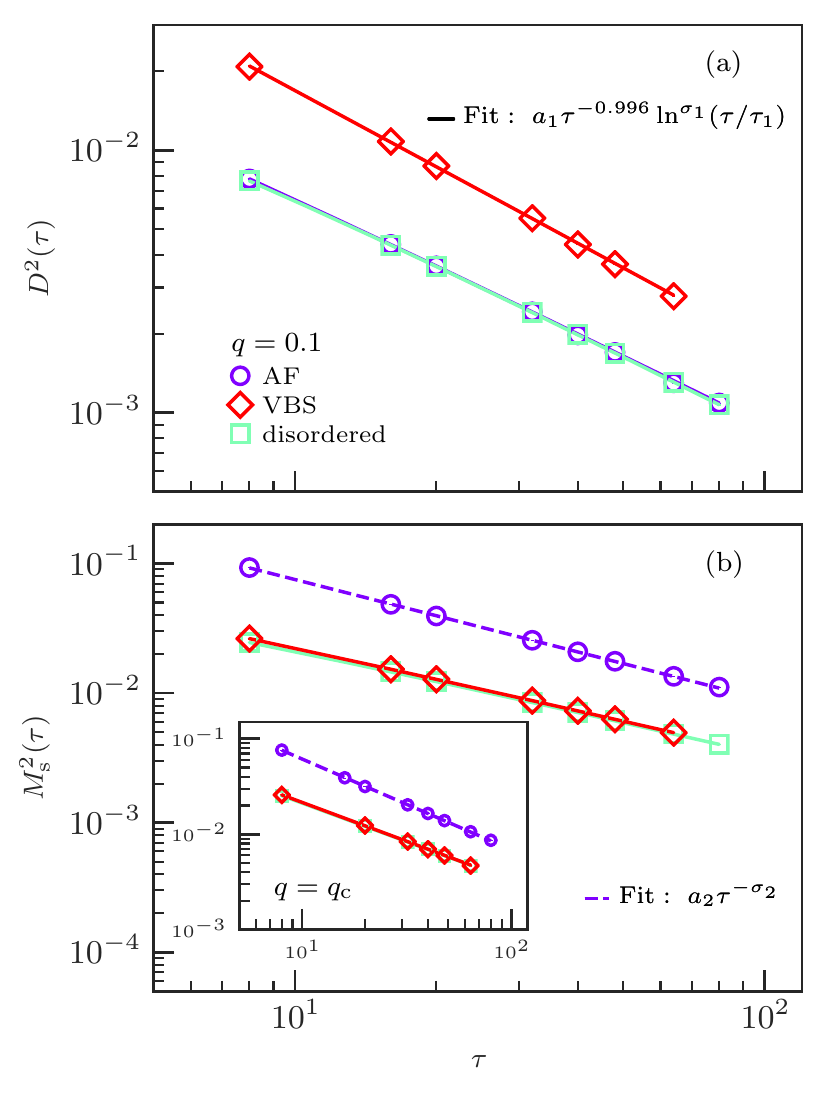}
  \vskip-3mm
  \caption{Dependence of $M_{\rm s}^{2}$ and $D^{2}$ on $\tau$ when relaxed from the AF/VBS/disordered state to $q=0.1$ and $q=q_{\rm c}$ with $\tau L^{-z}=1/16$. The solid/dashed lines correspond to a power-law form with/without logarithmic correction. The behaviors of $M_{\rm s}^{2}(\tau)$ and $D^{2}(\tau)$ are discussed in the text.}
  \label{fig:or_qlro}
\end{figure}

Since the perturbation which drives the dimerization transition in the J-Q$_3$ model is marginally irrelevant in the QLRO phase, we can infer that the scaling properties discussed above are also applicable in the QLRO phase up to a logarithmic correction~\cite{affleck1985prl,affleck1987prb}. To examine this, we perform QMC simulation with various initial states in the QLRO gapless phase. We find that for fixed $\tau L^{-z}=1/16$, $D^2(\tau)$ satisfies $D^2(\tau)\sim \tau^{-0.996}$ with a logarithmic correction as shown in Fig.~\ref{fig:or_qlro} (a). The fitting parameters are given by $a_{1}=0.1717(5)$, $\tau_{1}=6.8(6)$, $\sigma_{1} =0.021(5)$ for the VBS initial state, and $a_{1}=0.044(3)$, $\tau_{1}=1.0(2)$, $\sigma_{1} =0.44(3)$  for the disordered initial state and $a_{1}=0.021(6)$, $\tau_{1}=0.12(9)$, $\sigma_{1} =0.7(1)$ for the AF initial state.
The exponent therein is quite close to $2\beta/\nu z$ as at the critical point.

In addition, in Fig.~\ref{fig:or_qlro} (b), we also show the scaling behavior of the magnetic order parameter, which is defined as ${\bf M}_{\rm s}=(\sum_{i}^{L}{(-1)^{i}{\bf S}_{i})/L}$. We find that the squared staggered magnetization $M_{\rm s}^2(\tau)$ also obeys the scaling behavior $M_{\rm s}^2(\tau)\sim \tau^{-0.996}$ multiplied by a logarithmic correction term, for the VBS and disordered initial state. For the VBS case, we have $a_{1}=0.026(1)$, $\tau_{1}=0.04(3)$ and $\sigma_{1} = 1.2(2)$. For the disordered case, $a_{1}=0.024(7)$, $\tau_{1}=0.05(2)$ and $\sigma_{1} = 1.3(1)$. When the initial state has AF order, the logarithmic correction appears to be very weak and we instead use the pure power-law form in the fitting. We find the power is $\sigma_{2}=0.920(2)$ with $a_{2}=0.620(4)$, slightly different from the other two cases. This may because that the AF state is very far from the QLRO phase and the evolution time $\tau=L^{z}/16$ is so short. Whether there is a logarithmic correction in this situation needs more careful analysis.
In the inset of Fig.~\ref{fig:or_qlro} (b), we show $M_{\rm s}^2(\tau)$ at the critical point. One finds that they obey the same scaling behavior as in the QLRO phase. Here we list out the fitting parameter at $g=0$ for further reference. For the VBS case, $a_{1}=0.05(1)$, $\tau_{1}=0.11(6)$ and $\sigma_{1} =0.94(9)$. For the disordered case, $a_{1}=0.072(2)$, $\tau_{1}=0.3(1)$ and $\sigma_{1} =0.82(9)$. For the AF case, we use the pure power-law form, which gives $\sigma_{2}=0.938(1)$ and $a_{2}=0.525(2)$.
As discussed above, the logarithmic corrections found here could be induced by short-time scaling corrections or inaccurate value of critical point.
Finding out the origin of the logarithmic corrections is beyond the purpose of this study and we will leave it to further studies.

\section{summary}
\label{sec:conclusion}

In this work, we have studied the SITQCD of the QLRO-VBS transition in the J-Q$_{3}$ chain. Using the method based on the scaling theory of the SITQCD, we have determined its critical point as $q_{\rm c}=0.170(14)$, in agreement with a recent ED and QMC study~\cite{yang2020}. Then we have determined the critical initial slip exponent $\theta=-0.507(3)$ and the static exponent $\beta/\nu=0.498(2)$. Moreover, by comparing the value of $\theta$ and $\beta/\nu$, we have found that the initial order parameter $D_0$ is a marginal scaling variable. This is quite different from the case in the quantum Ising model, in which the initial order parameter is a relevant scaling variable~\cite{yin2014prb,shu2017prb}. We have shown that the marginal $D_0$ leads a short-time decay of the order parameter, rather than the initial increase as shown in the quantum Ising model~\cite{yin2014prb,shu2017prb}. We also have argued that the reason for the appearance of the marginal initial order parameter is that this phase transition is induced by a perturbation, which is marginally irrelevant in the QLRO phase and marginally relevant in the VBS phase~\cite{affleck1985prl,affleck1987prb}. Accordingly, we have also shown that the scaling theory of the SITQCD at the critical point is also applicable in the QLRO phase, only up to a logarithmic scaling correction.

Recently, the critical initial slip behavior was also found theoretically in the prethermal real-time dynamics~\cite{gagel2014prl,gagel2015prb,calabrese2012jsm_1,calabrese2012jsm_2,maraga2015pre,chiocchetta2015prb,chiocchetta2016prb_1,chiocchetta2016prb_2,chiocchetta2017prl,liu_w2016jpa}. In particular, a negative initial slip exponent was also found in the quench dynamics of the Dirac systems~\cite{jian2019prl}. Accordingly, it is instructive to study the real-time relaxation dynamics of J-Q$_{3}$ model, which we leave as a further work.
Besides, due to similarity between the imaginary-time relaxation and boundary effect in real space, it is also interesting to consider the effect of a marginal $D_{0}$ in real space~\cite{diehl1986phase}. In a system with the boundary set to have a fixed local $D_{0}$, the dependence of $D$ on the distance to the boundary $r$ should obey $D(r)\sim r^{-\beta/\nu}$ if $D_{0}$ is marginal. The effect of $D_{0}$ is not propagated through the space due to its marginal role. This issue is also worth investigating.
More interestingly, the two-dimensional J-Q$_{3}$ model hosts a N{\'e}el-VBS quantum phase transition beyond the Landau-Ginzburg-Wilson (LGW) paradigm~\cite{senthil2004sci,sandvik2007prl}. The SITQCD has proved applicable in LGW phase transitions as well as topological quantum phase transitions~\cite{yin2014prb}. The SITQCD in the deconfined quantum phase transition framework is very intriguing and this work is in process.

\acknowledgments{The authors are grateful to Anders W. Sandvik for his valuable discussions. Y.-R.S. acknowledges support from Grant No. NSFC-11947035 and the startup grant (RP2020120) at Guangzhou University. S.Y. is supported by the startup grant (No. 74130-18841229) at Sun Yat-Sen University.}

\bibliography{ref}
\end{document}